\documentclass[%
 reprint,
 amsmath,amssymb,
 aps,
]{revtex4-2}

\usepackage{hyperref,color}

\usepackage{graphicx}
\usepackage{dcolumn}
\usepackage{bm}

\usepackage{fullwidth}

\begin{document}

\preprint{APS/123-QED}

\title{Modeling the Transition between Localized and Extended Deposition in Flow Networks through Packings of Glass Beads}

\author{Gess Kelly$^{1}$}
 \author{Navid Bizmark$^{3,4}$}
  \author{Bulbul Chakraborty$^{1}$}
 \author{Sujit S. Datta$^{3}$}
 \author{Thomas G. Fai$^{2}$}%
  \email{tfai@brandeis.edu}
\affiliation{%
$^{1}$Martin A. Fisher School of Physics, Brandeis University, Waltham, MA 02453
}%
\affiliation{%
$^{2}$Mathematics Department \& Volen Center for Complex Systems, Brandeis University, Waltham, MA 02453
}%
\affiliation{
$^{3}$Department of Chemical and Biological Engineering, Princeton University, Princeton, NJ 08544, USA
}
\affiliation{
$^{4}$Princeton Materials Institute, Princeton University, Princeton, NJ 08540, USA
}
%



\date{\today}

\begin{abstract}
We use a theoretical model to explore how fluid dynamics, in particular, the pressure gradient and wall shear stress in a channel, affect the deposition of particles flowing in a microfluidic network.  Experiments on transport of colloidal particles in pressure-driven systems of packed beads have shown that at lower pressure drop, particles deposit locally at the inlet, while at higher pressure drop, they deposit uniformly along the direction of flow. We develop a mathematical model and use agent-based simulations to capture these essential qualitative features observed in experiments. We explore the deposition profile over a two-dimensional phase diagram defined in terms of the pressure and shear stress threshold, and show that two distinct phases exist. We explain this apparent phase transition by drawing an analogy to simple one-dimensional models of aggregation in which the phase transition is calculated analytically.
\end{abstract}

\maketitle



\section{\label{sec:level1}Introduction}
Deposition and aggregation of fine particles in microfluidic networks and porous media play an important role in various natural and industrial processes such as water purification, geotextile filtration, applications in precision drug delivery and similar biomedical tasks, transport of microplastics, environmental cleanups, groundwater pollutant removal, {oil recovery,} and transport of nanomaterials for groundwater aquifer remediation \cite{LinkhorstJohn2016Mcf, FaureY.H2006Acfp, drugKLEINSTREUER20085590, microplasticAlHarraqAhmed2022MttL, ZHAO2016245,kanel,phenrat2009} {\cite{BoekMeso2008,BoekMulti2010,oilRec1Boek2010,Lawal2012}}. For example, in filtration processes, understanding of the deposition dynamics of colloidal particles plays a significant role in improving filter efficiency via reducing filter fouling  \cite{colloidalaggregates, membranefouling, M.P.Dalwadi2015Uhpg}. Observations from \cite{BizmarkNavid2020Mdoc} indicate that, regardless of the charge of the colloidal particles flowing in the bead network, applying lower pressures across the system leads to localized deposition under various conditions. This may suggest that irrespective of the exact local clogging mechanism (e.g., bridging versus aggregation \cite{DressaireEmilie2016Coms}), the interplay of hydrodynamical variables in these systems controls the resulting deposition profile. We focus on the role of applied pressure difference $\Delta P$ as one of the key variables motivated by the experimental design in \cite{BizmarkNavid2020Mdoc} and the wall shear stress $\tau_w$, which has been shown in past studies to play an important role in erosion \cite{shearRistrophLeif2012Soae, shearSCHORGHOFERNORBERT2004Scip, JagerR2017Cipm}. Here, the shear stress at the wall $\tau_w$ refers to the shear stress experienced by colloidal particles deposited on the walls of the packing. We follow the approach of \cite{JagerR2017Cipm} to capture the role of the shear stress threshold $\tau$, a material parameter that describes the threshold shear stress at the wall above which fluid flow erodes the deposited particles from the walls. {Table \ref{tab:variables} in the Supplementary Material contains representative parameter values.}  Throughout the text, we use a hat notation, e.g., $\Delta \hat{P}$, to denote the corresponding variables, e.g., $\Delta P$, that are normalized by a set value relevant to the experimental system. Table \ref{Tab:2} in the Supplementary Material contains additional details. 


Our specific system of interest is motivated by recent experiments from \cite{BizmarkNavid2020Mdoc}, in which a constant pressure difference $\Delta P$ applied to a packing of disordered glass beads of length $L$ drives a fluid flow containing a suspension of colloidal particles. These experiments show that at larger pressure differences, the profile of particles deposited on the solid matrix extends uniformly along the length of the packing, while at lower pressures, the particles deposit locally at the inlet where they are injected into the system. Here, we develop a mathematical model to explain how the pressure difference influences the deposition profile.

\begin{figure}
    \centering
    \includegraphics[width = 9cm]{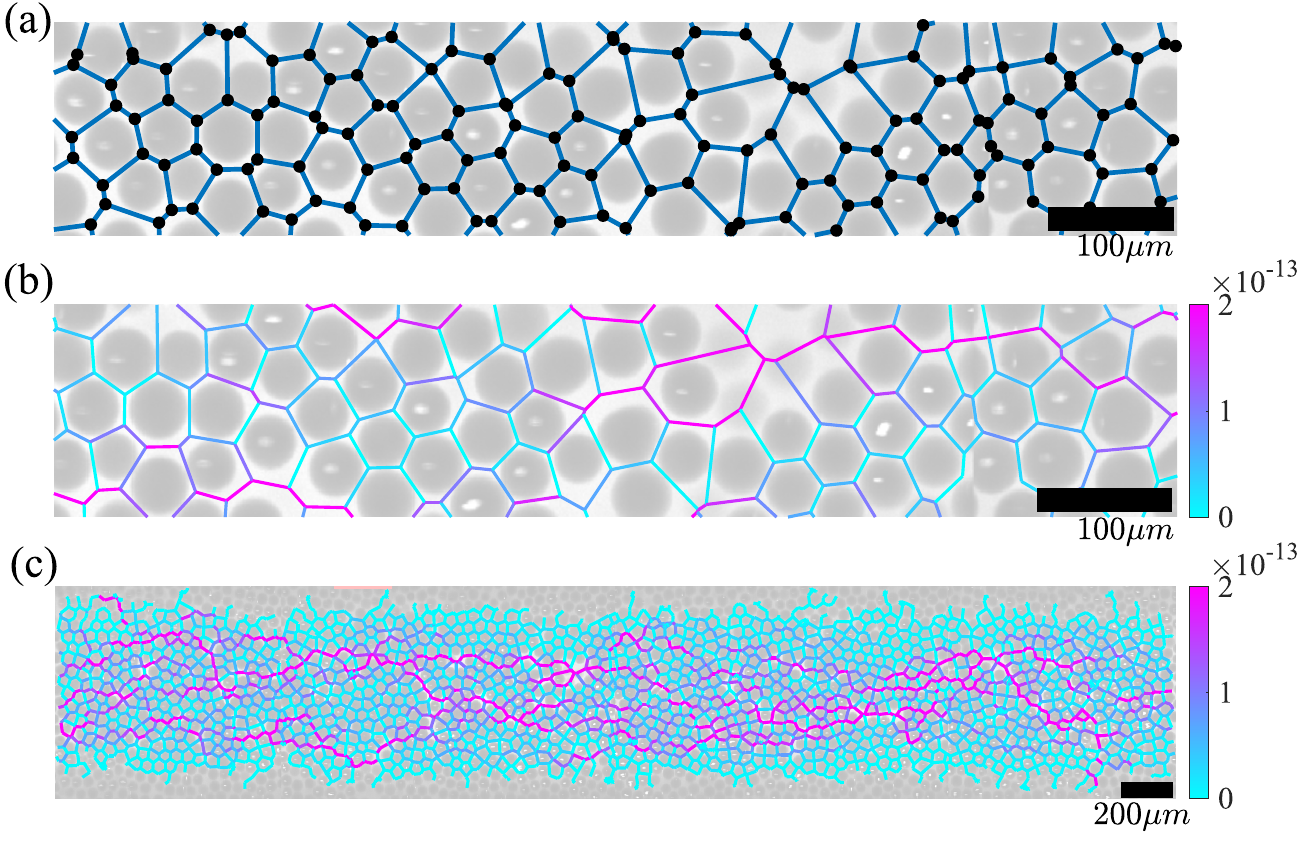}
    \caption{We use a network approach to model the bead packing here shown in the absence of particles. (a) We skeletonize the image of the packing, and then generate a network. The edges of the network represent the channels through which fluid may flow in the packing and the nodes represent the junctions where these channels meet.  (b) We obtain the flow rates in the channels by applying the Kirchhoff's laws \cite{Bruus}. (c) Zoomed-out view showing the network as a whole. The color in (b) and (c) shows the magnitude of the channel flow rates in SI units $(m^3/s)$. (a) and (b) have the same scale bar. The grey background shows the experimental micrograph of the beads. }
    \label{fig:graph}
\end{figure}

Past studies of simple mass-aggregation models \cite{MajumdarSatyaN1998Npti, M.P.Dalwadi2015Uhpg} motivate us to explore the phase space of shear stress threshold ${\hat{\tau}}$ and pressure difference $\Delta \hat{P}$. In particular, Majumdar et al.~\cite{MajumdarSatyaN1998Npti} consider minimal systems and lattice models in which discrete masses diffuse at a constant unit rate, which normalizes the overall timescale. Multiple masses may aggregate at lattice sites after diffusion, and units of masses erode (chip away) from blocks at a constant chipping rate $w$. Physically, chipping corresponds to single-particle dissociation in processes such as polymerization and competes with coalescence.  In this simplest case, they work with two independent variables, the chipping rate $w$ and mass density $\rho$, that remain constant and determine the behavior of the system at steady state. They explore the phase space consisting of the mass density $\rho$ and chipping rate $w$ and show that these finite systems exhibit two distinct phases at steady state, only one of which involves an infinite aggregate. One important difference between the simple mass-aggregation model and our study is the fixed density or constant total mass with periodic boundary conditions in contrast to our model where there is a flux of particles into and out of the system.

We formulate the fluid flow through the packings by applying the hydraulic analogy to the network of channels extracted from the bead packing images. Using our network model and deposition and erosion laws, we demonstrate a similar transition in the normalized shear stress threshold $\hat{\tau}$ and pressure $\Delta \hat{P}$ phase space.  Motivated by these simple models of aggregation and fragmentation explored in previous studies \cite{RajeshR2001Epdo, MajumdarSatyaN1998Npti}, we explore the model phase space spanned by two dimensionless parameters, and identify a transition between extended and localized deposition regimes in terms of the key parameters of pressure difference and shear stress threshold \footnote{There are different combinations of parameters (such as flow rate and pressure) that one could potentially choose, but as long as there is a way to map from one to the other, they are expected to be equivalent.}.


\section{Methods}

We use a graph- or network-based approach \cite{bollobas_graph, barrat_ntwrk_2008} to model the porous network created by the beads as shown in FIG. \ref{fig:graph}(a). The idea of modeling a porous system as a network has been studied previously \cite{GriffithsI.M2014Acnm, bubb_Redner2011, ZareeiAhmad2021TEoF}. For instance, past studies have demonstrated the effectiveness of a network-based approach by highlighting the role of disorder on the flow distribution in porous media \cite{AlimKaren2017LPSC}. We use images of two-dimensional (2D) slices of the three-dimensional (3D) packing. We then generate the model network based on these images. Because of the expected differences between the flow in 2D and 3D, we do not expect to quantitatively recover all aspects of the experiments.  In such network models, each pore or channel is typically represented by an edge in a network representing the entire porous system (see FIG. \ref{fig:graph}.(a)). Each edge may be weighted in terms of its conductance and the nodes of the network represent junctions between the edges. Assuming we have an incompressible fluid, the inflow and outflow of particles and fluid must be equal to respect mass conservation. In our system of interest, boundary junctions at the inlet and outlet are subject to two pressures held constant for the duration of the experiment. To solve for the resulting channel flow rates, as shown in FIG. \ref{fig:graph} (b) and (c), we apply Kirchhoff's circuit laws. For each channel, we estimate the channel length $l$ and diameter $d$ from the image of the network to calculate the channel conductance $g$, which is defined as the proportionality constant between the volumetric flow rate through a given channel and the pressure difference across the channel given by the Hagen-Poiseuille law {\cite{Bruus}}:
\begin{equation}
    g = \frac{\pi d^4}{128 \eta l},
\end{equation}
where $\eta $ is the dynamic viscosity.  The resolution of the image in FIG. \ref{fig:graph} tends to be lower along the boundaries and our image processing does not accurately identify a significant portion of the channels. We use the largest connected component of the model network, which is in the interior of the packing. For this reason we neglect the upper and lower boundaries. More details regarding channel flow rate calculations can be found in the Supplementary Material. The total flow rate is of order $10^{-10}\, m^3/s$ once we account for the depth of the three-dimensional system. 

To capture the stochastic effects, we use agent-based modeling to model the particles as they deposit and erode within the network. This distinguishes our study from a closely related previous model of erosion in networks in which differential equations are used to predict how erosion changes the width of the channels in the network \cite{ZareeiAhmad2021TEoF}. Another difference is our assumption that the glass beads that form the network remain fixed over the course of the simulation. Consequently, while the deposited particles may erode in our simulation, the channels themselves do not erode. {Because initially the channel does not contain any deposited particles, and since erosion only occurs through removal of particles, the channel width cannot grow beyond its initial value.}

Particles enter the system from the inlet at constant time intervals. This is a discrete approximation to the experiments, in which the particles are injected continuously at a constant volume fraction. This is also different from the conserved-mass aggregation models of \cite{MajumdarSatyaN1998Npti} where the density is constant. As particles deposit in the network during the simulations, they cause a decrease in the width of the channels, which may eventually lead to topological changes when the number of deposited particles surpasses the channel capacity, i.e., clogging. We assume that each time a particle is deposited {(eroded)}, it causes a uniform reduction {(increase)} in the channel width. This assumption is motivated by the separation of length scales in the experiments, in which the glass bead diameter is approximately $40 \mu m$ so that the particle-to-bead size ratio is approximately $0.03$.   

We follow the suggested model of \cite{JagerR2017Cipm} in defining the deposition rate $\lambda_d$ and erosion rate $\lambda_e$ of particles using shear stress thresholds: the deposition threshold $\tau_d$ and erosion threshold $\tau_e$, and shear stress at the channel wall $\tau_w$. Since we are interested in cases in which both deposition and erosion occur, to reduce the number of independent parameters, here, we assume that the wall shear stress thresholds for deposition and erosion are equal, i.e., $\tau_d = \tau_e = \tau$, so that the deposition rate and erosion rate equations are: 
  \begin{equation}
  \label{eq:dep}
    \lambda_d(\tau) =
    \begin{cases}
      \kappa_d (\tau - \tau_w), & \text{if}\ \tau_w < \tau \\
      0, & \text{otherwise,}
    \end{cases}
  \end{equation}
  
  and 
  
  \begin{equation}
  \label{eq:er}
    \lambda_e(\tau) =
    \begin{cases}
      \kappa_e (\tau_w - \tau), & \text{if}\ \tau_w > \tau \\
      0, & \text{otherwise.}
    \end{cases}
  \end{equation}

Here, $\kappa_d$ and $\kappa_e$ are deposition and erosion coefficients that depend on solid properties, respectively \cite{JagerR2017Cipm}. We note that $\tau_w$ depends on the imposed fluid flow conditions, whereas $\tau_d$ and $\tau_e$ reflect the material properties of the deposited particles independent of flow. In particular, a larger $\tau$ requires a larger wall shear stress for particle erosion. We run the simulation for multiple values of $\Delta \hat{P}$ and $\hat{\tau}$, keeping all other parameters, including the length of the medium $\hat{L}$, constant.

 \begin{figure}
    \centering
    \includegraphics[width = 8.5cm]{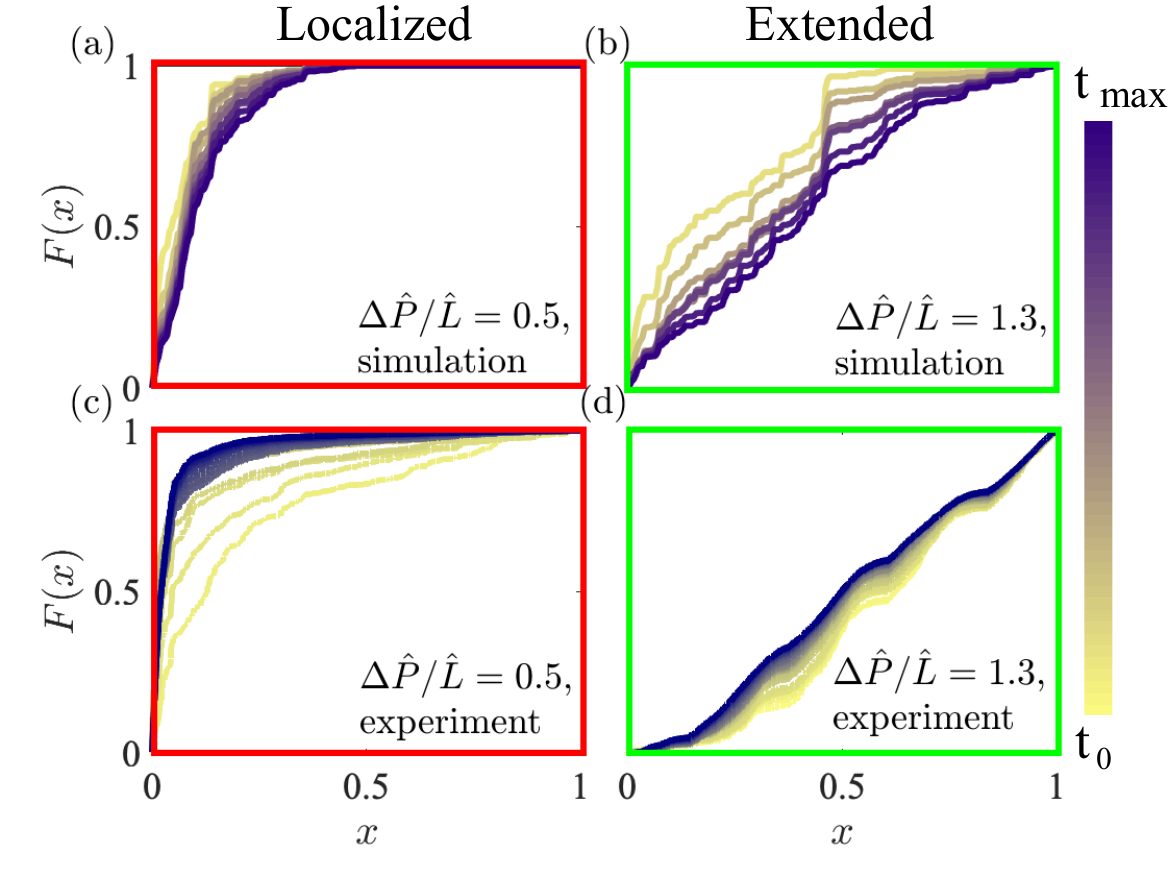}
    \caption{The cumulative probability distribution function $F(x)$ of positions of particles deposited along the flow direction for the localized and the extended case obtained by simulation and experiment show a similar qualitative behavior. Darker colors indicate later times. The position along the direction of the flow $x$ is normalized by the total length of the medium. The labels indicate the normalized pressure gradient.}
    \label{fig:Fx}
\end{figure}

\begin{figure*}
    \centering
    \includegraphics[width = 17.5cm]{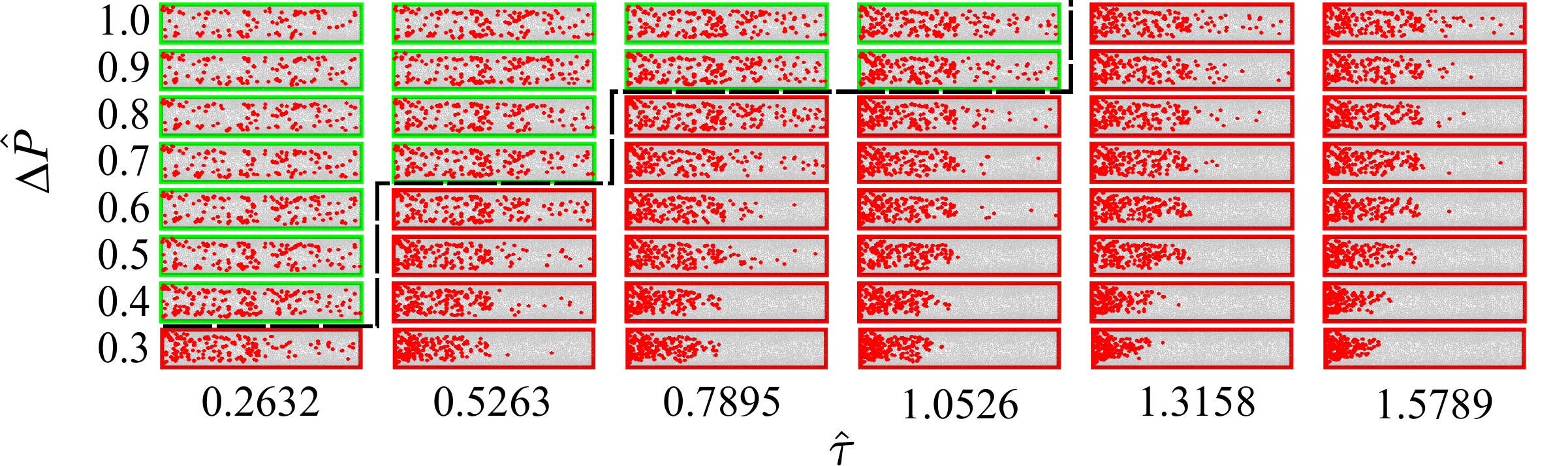}
\vspace{-.5cm}
    \caption{The final frames of the simulation over a range of values of shear stress threshold $\hat{\tau}$ and applied pressure $\Delta \hat{P}$ show a clear separation between the localized and extended deposition regimes similar to experimental observations in \cite{BizmarkNavid2020Mdoc}. The dashed line serves to guide the eye. }
    \label{fig:recs}
\end{figure*}

\begin{figure}
    \centering
    \includegraphics[width = 8.5cm]{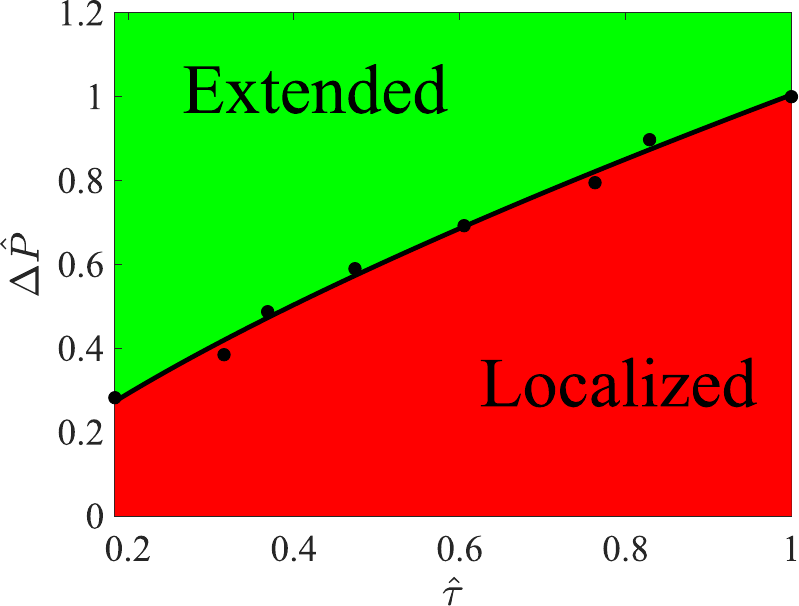}
    \caption{In the parameter space of normalized pressure $\Delta\hat{ P}$ and shear stress threshold $\hat{\tau}$, a boundary separates the two regimes of localized and extended deposition, reminiscent of the phases in \cite{MajumdarSatyaN1998Npti}. The filled circles show the critical values of pressure $P_c$ at which the transition to the localized phase occurs in simulations. The solid line corresponds to the best fit {$\Delta\hat{ P}={1.04\hat{\tau}^{0.71}-0.04}$}. The hat notation used here denotes normalization by the maximum value.}
    \label{fig:phasecurve}
\end{figure}
 
\section{Results}
In both experiments and simulations, the cumulative distribution function $F(x)$ of deposited particle positions varies significantly between the localized and extended deposition regimes.
{We approximate this function as
\begin{equation}
    F(x) \approx N(X \leq x)/N, 
\end{equation}
where $N(X \leq x)$ is the number of deposited particles at position less than or equal to $x$, and $N$ is the the total number of deposited particles. The position along the direction of the flow $x$ is normalized by the full length of the medium.} As shown in FIG. \ref{fig:Fx}, when the deposition is localized, $F(x)$ attains a value near 1 for $x < 0.5$, showing that most of the deposited particles are close to the inlet. In contrast, in the extended deposition case,  $F(x)$ has a more linear form with $F(x) \approx 0.5 $ when $ x = 0.5 $. More information regarding the normalization constants is included in Table \ref{Tab:2} in Supplementary Material.

Our simulations reveal that for each wall shear stress threshold $\hat{\tau}$, there exists a critical pressure $\Delta \hat{P}_c$ that separates the localized and extended regimes (FIG. \ref{fig:recs}). For a specific value of  $\hat{\tau}$, choosing $\Delta \hat{P}$ larger {(smaller)} than the critical value leads to extended {(localized)} deposition. To find the critical pressure $\Delta \hat{P}_c$ for a given $\hat{\tau}$, we vary $\Delta \hat{P}$ in the simulation while keeping all other parameters constant. As we decrease $\Delta \hat{P}$, the percentage of the deposited particles in the first half of the system increases. We mark $\Delta \hat{P}_c$ as the smallest $\Delta \hat{P}$ when the deposition is localized. FIG. \ref{fig:phasecurve} visualizes $\Delta \hat{P}_c$ for various values of $\hat{\tau}$ and how the two regimes of localized and extended deposition are separated in the normalized shear stress threshold and pressure phase space. This behavior is similar to the phase transitions observed in simple mass-aggregation models on lattice sites, with shear stress and pressure appearing on the corresponding axes of the phase diagram as the chipping rate and density in the chipping and aggregation model \cite{MajumdarSatyaN1998Npti,MajumdarSatyaN2000NPTi}.  {The solid curve in FIG. \ref{fig:phasecurve} is the best power law fit motivated by the power law relation between the model parameters in \cite{MajumdarSatyaN1998Npti}. In particular, we fit to a function of the form $ax^b+c$. }

Majumdar et al. use the steady-state mass distribution to study the behavior and dynamical phase transition of their model, in which the distribution transitions from an exponential to a power-law with an aggregate \cite{MajumdarSatyaN1998Npti}. Our numerical investigation of the mass distribution has revealed signs of a similar behavior when transitioning between the extended and localized phases. However, an accurate classification of the transition seen in our model requires a more rigorous study of the critical point. Identifying the transition point in complex non-equilibrium systems such as ours where a free energy description of the system does not exist is a difficult task, and as a first step, one may simplify some of the complexities of the system to map it to other solvable models. 

In the experiments, one of the tunable parameters is $\Delta \hat{P}$. We further assume that $\hat{\tau}$ is an independent parameter that depends on fluid, particle, and pore network properties. Given a system with a fixed $\hat{\tau}$, we expect localized deposition at lower $\Delta \hat{P}$, and extended deposition at higher $\Delta \hat{P}$ for the same network as seen in experiments \cite{BizmarkNavid2020Mdoc} and simulations. FIG. \ref{fig:recs} supports this reasoning. For particles with a given shear stress threshold $\hat{\tau}$, as we increase $\Delta \hat{P}$, a smaller percentage of deposition occurs near the inlet of the medium, consistent with the experimental findings. In the figure, the red borders indicate localized deposition, as defined by comparison to a representative experiment, and the green borders indicate extended deposition. More details on categorizing localized and extended deposition are found in the Supplementary Material. In the SM, we also demonstrate that this transition occurs over a range of sizes. The hat notation used here denotes normalization by a set value.  The programming scripts used in generating the simulations discussed in this section are accessible on GitHub \cite{kellyGitHub}.               



\section{Discussion}
Capturing the dynamics of deposition in porous media and microfluidic systems has wide implications in filtration studies. Understanding what leads to localized deposition helps in improving filter efficiency \cite{GriffithsI.M2014Acnm}. Experiments such as \cite{BizmarkNavid2020Mdoc} provide more insight into the influence of global system hydrodynamics on uniformity of deposition profiles. Our theoretical model successfully captures the behavior observed in the experiments in \cite{BizmarkNavid2020Mdoc}. Our network-based approach and model of shear-based deposition and erosion reveal a transition from the localized to extended regime in the phase space of shear stress threshold and pressure in colloidal transport within packings of beads. Given a system of beads, there exists a critical pressure above which the deposition profile becomes increasingly more uniform (see FIG. \ref{fig:phasecurve}). This transition from the localized to extended regime is similar to what has been observed in previous studies of simple mass-aggregation models \cite{MajumdarSatyaN1998Npti}. The observation in similarities between these models of aggregation and our current model leads us to believe such analogies may be present in other systems, as well, where the key variables may be different. Future applications to other systems including filtration may examine and identify what variables control the phase transition in the system. 

Our model may be thought of as a two-dimensional inhomogeneous asymmetric expansion of the mass-aggregation model in \cite{MajumdarSatyaN1998Npti}. Uncovering the limitations to this analogy requires a careful analysis. One important distinction between the two models concerns the boundary conditions. In the simple mass-aggregation model, the total mass is constant, whereas in our model, there is a regular influx of particles into the system and particles may exit at the outlet boundary. Some other relevant models that allow an influx of particles include the aggregation-chipping model with open boundary conditions \cite{himanibarma2013} and totally asymmetric simple exclusion (or inclusion) processes on networks \cite{NetworkTASEPNeri, ReuveniAsip2014}. In our case, the direction of flow makes the model asymmetric, and although the simple mass-aggregation model also shows a phase transition in two dimensions, it belongs to a different universality class \cite{MajumdarSatyaN2000NPTi,RajeshKrishnamurthyBias}. Moreover, Rajesh et al.~have shown that subtle changes such as making the deposition rate mass-dependent lead to different models with no phase transitions \cite{RajeshR2002Afia}. In our case, we assume identical particles with the same laws for deposition and erosion that do not explicitly depend on mass of individual particles; however, the flow, deposition, and erosion rates may change with time and differ for each channel. These differences lead to spatial bias and signs of channelization \cite{ZareeiAhmad2021TEoF, Kudrolli2016, JagerR2017Cipm} where particles frequent a few paths rather than all paths in the system. Studies suggest in real world applications of the model, the rate of erosion and shear threshold for erosion may depend on particle-particle interactions and spatial distribution of particles \cite{Kudrolli2016}, similar to a mass-dependent law. {Past studies considering the role of particle interactions in deposition report that strong particle interactions lead to a decrease in the transient flow rate \cite{BoekMeso2008}. We hypothesize that this decrease would lead to a lower effective local shear stress at the channel walls and hence a more localized deposition profile.} Since particles act as agents in the model and simulations, {adding interactions} would be a possible expansion of the model in the future.

We note that, although the model is successful in capturing the essential behavior of the system, some details regarding the clogging mechanism are lost due to coarse-graining. This is most apparent in FIG. \ref{fig:Fx} in which the deposition appears to become more localized over time in experiments in contrast to simulations. It would be interesting to explore what leads to this difference in experiment and simulation observations by expanding the model to three dimensions. Additionally, one may expand \eqref{eq:dep} and \eqref{eq:er} to consider an \textit{overlap} region such that $\tau_e < \tau_d$ where for some values of wall shear stress, both deposition and erosion occur, or a \textit{gap} region $\tau_d < \tau_e$ where for some values of wall shear stress, neither deposition nor erosion occur similar to the generalizations in \cite{JagerR2017Cipm}.  


\section{Acknowledgements}
We acknowledge funding from NSF grant DMS-1913093 to GK and TGF. We also acknowledge computational support from the Brandeis HPCC which is partially supported by the NSF through DMR-MRSEC 2011846 and OAC-1920147. SSD acknowledges funding from NSF DMR-2011750. NB acknowledges support from a postdoctoral fellowship of the Princeton Center for Complex Materials (PCCM). BC acknowledges the support of NSF-CBET 1916877. 


\bibliography{ref}

\newpage
\section{Supplementary Material}

\renewcommand{\thefigure}{S\arabic{figure}}
\renewcommand{\thetable}{S\arabic{table}}
\renewcommand{\theequation}{S\arabic{equation}}

\subsection*{S1. Solving for the Channel Flow Rates}
\label{s1:flow_rate}
To solve for the channel flow rates in the system, we first start by processing the image of the glass bead packing and formulating the corresponding network.  Then, following the steps outlined in \cite{resNetwork2010} closely, we solve for the local pressure and channel fluid dynamics. 

Given an image of the packing of glass beads, we use built-in functions in MATLAB to binarize and skeletonize the image \cite{ZHANG20041, CHATBRI201659, JiangZ2007Eeon, MATLAB}. Then, we use the package Skel2Graph3D \cite{KollmannsbergerPhilip2017Tswo} to generate the graph (network) $\mathbf{G}$ representing the skeleton network. Each edge in the network represents a channel (pore) and stores its geometrical properties such as its diameter $d_j$ and length $l_j$, which are calculated from the image, and its hydrodynamics such as its flow rate $q_j$, shear stress at the wall $\tau_j$, and conductance $g_j$. Each node $n_i$ in the network represents a junction where multiple channels meet and stores its position $\mathbf{x_i} $ and pressure $p_i$. Similar to the experiments, the pressure values at the boundary nodes at the entrance and exit of the flow in the system are prescribed and kept constant throughout the simulation. This allows us to solve for the unique values of pressure at the bulk nodes using the Kirchhoff's laws.     

To solve for the pressure in the bulk, first, we rearrange the pressure $\mathbf{p}$ and nodal flow rate values $\mathbf{j}$, separating the ones associated with the boundary and bulk, and ordering them such that the boundary values denoted as $j_B$ and $p_B$ appear at the top of the vectors and the internal connection values follow after:
\begin{equation}
    \mathbf{j} = \begin{bmatrix}
\mathbf{j_B} \\
0 
\end{bmatrix},
\end{equation}
and 
\begin{equation}
    \mathbf{p} = \begin{bmatrix}
\mathbf{p_B} \\
\mathbf{p_C} 
\end{bmatrix}.
\end{equation}
Note that since this is a boundary value classical Kirchhoff's problem with fixed pressure at the boundary nodes, the nodal flow rate at the bulk (non-boundary nodes) $\mathbf{j_C} = 0$ due to conservation of mass. We then formulate the weighted Laplacian $\mathcal{L}$ of the porous structure graph (network) 
\begin{equation}
    \mathcal{L} = \mathcal{D} \mathbf{g} \mathcal{D}^T,  
\end{equation}
where $\mathcal{D}$ denotes the incidence matrix, $\mathcal{D}^T$ its transpose, and $\mathbf{g} $ the vector of channel conductances. The solution to the boundary nodal flow rates is given by 
\begin{equation}
    \mathbf{j_B} = \mathcal{L}_S \mathbf{p_B},  
\end{equation}
where $\mathcal{L}_S$ is the Schur complement of the Laplacian matrix. To obtain the bulk pressures $\mathbf{p_C} $, we then solve the inverse matrix problem
\begin{equation}
    \begin{bmatrix}
\mathbf{j_B} \\
0 
\end{bmatrix} = \mathcal{L} \begin{bmatrix}
\mathbf{p_B} \\
\mathbf{p_C} 
\end{bmatrix}.  
\end{equation}

By Ohm's law, to obtain channel flow rate $q_j$ for channel $j$, we use the pressure difference at the channel end-nodes, $p_{j1}, p_{j2}$:
\begin{equation}
    q_j = g_j (p_{j1} - p_{j2}),
\end{equation}
where we calculate the conductance $g_j$ by approximating the channels as cylinders with diameter $d_j$ and length $l_j$:
\begin{equation}
    g_j = \frac{\pi d_j^4}{128 \eta l_j}.
\end{equation}
We may now also calculate the channel shear stress $\tau_j$:
\begin{equation}
    \tau_j = \frac{32 q_j \eta}{\pi d_j^3}.
\end{equation}

\subsection*{S2. Modeling the Particle Deposition and Erosion}
\label{s2:deposition}
Particles are injected at regular time steps. The injection period, $T_\text{inj}$ the time passed in between each two consecutive particle injections, is calculated as follows
\begin{equation}
    T_\text{inj} = 2 T_\text{transit},
\end{equation}
where $T_\text{transit}$ is the approximate time that it takes the flow to take one particle from the inlet to the outlet:
\begin{equation}
    T_\text{transit} = \frac{L}{\overline{q}/\overline{d}^2}.
\end{equation}
Here, $L$, $\overline{q}$, and $\overline{d}$ are the total length of the medium, the mean channel flow rate and the mean channel diameter.
Each particle is initialized as an object at a random inlet boundary node and stores its position (floating point), edge number (integer), node number (integer), and whether it is deposited (logical). The node number indicates if a particle has arrived at a junction where the number corresponds to the node and is otherwise set to 0. When a particle arrives at a node, it is assigned an edge that connects that node to a node with lower pressure. If there are multiple such edges, the edge is picked with a probability that is proportional to the flow rate in the edge. The edge number of the particle is then set to reflect its new edge assignment. After the particle is assigned an edge, it travels with the channel speed $v_j$ calculated from the flow rate:
\begin{equation}
    v_j = 4 q_j/(\pi d_j^2).
\end{equation}
Choosing an appropriately small time step is crucial here, since higher pressure systems have higher mean flow rates and thus higher transport velocities. The time step $\Delta t$ must be small enough to get accurate results and to avoid unintentionally large transport distances in a channel. While traveling in an edge if the channel shear stress $\tau_j= \tau_w$ is lower than the threshold for deposition $\tau_d$, the particle may deposit with the probability $\theta_d$ at each time step $\Delta t$:
\begin{equation}
    \theta_d = \kappa_d (\tau_d - \tau_w) \Delta t,
\end{equation}
where $\kappa_d $ is the deposition coefficient. If the channel shear stress is above this threshold, $\theta_d = 0$. If the particle deposits, its deposition number is changed from 0 to 1. At each time step, if the channel shear stress $\tau_j = \tau_w$ is higher than the threshold for erosion $\tau_e$, a deposited particle may erode with the probability  
\begin{equation}
    \theta_e = \kappa_e (\tau_w - \tau_e) \Delta t,
\end{equation}
where $\kappa_e$ is the erosion coefficient. The capacity of each edge is determined by the ratio of its diameter and the diameter of the particles. When $\tau_e = \tau_d = \tau$,

  \begin{equation}
    \theta_d(\tau) =
    \begin{cases}
      \kappa_d (\tau - \tau_w) \Delta t, & \text{if}\ \tau_w < \tau \\
      0, & \text{otherwise}
    \end{cases}
  \end{equation}

and

\begin{equation}
    \theta_e(\tau) =
    \begin{cases}
      \kappa_e (\tau_w - \tau) \Delta t, & \text{if}\ \tau_w > \tau \\
      0, & \text{otherwise.}
    \end{cases}
  \end{equation}

\begin{figure}
    \centering
    \hspace{-.2in}\includegraphics[width = 0.5\textwidth]{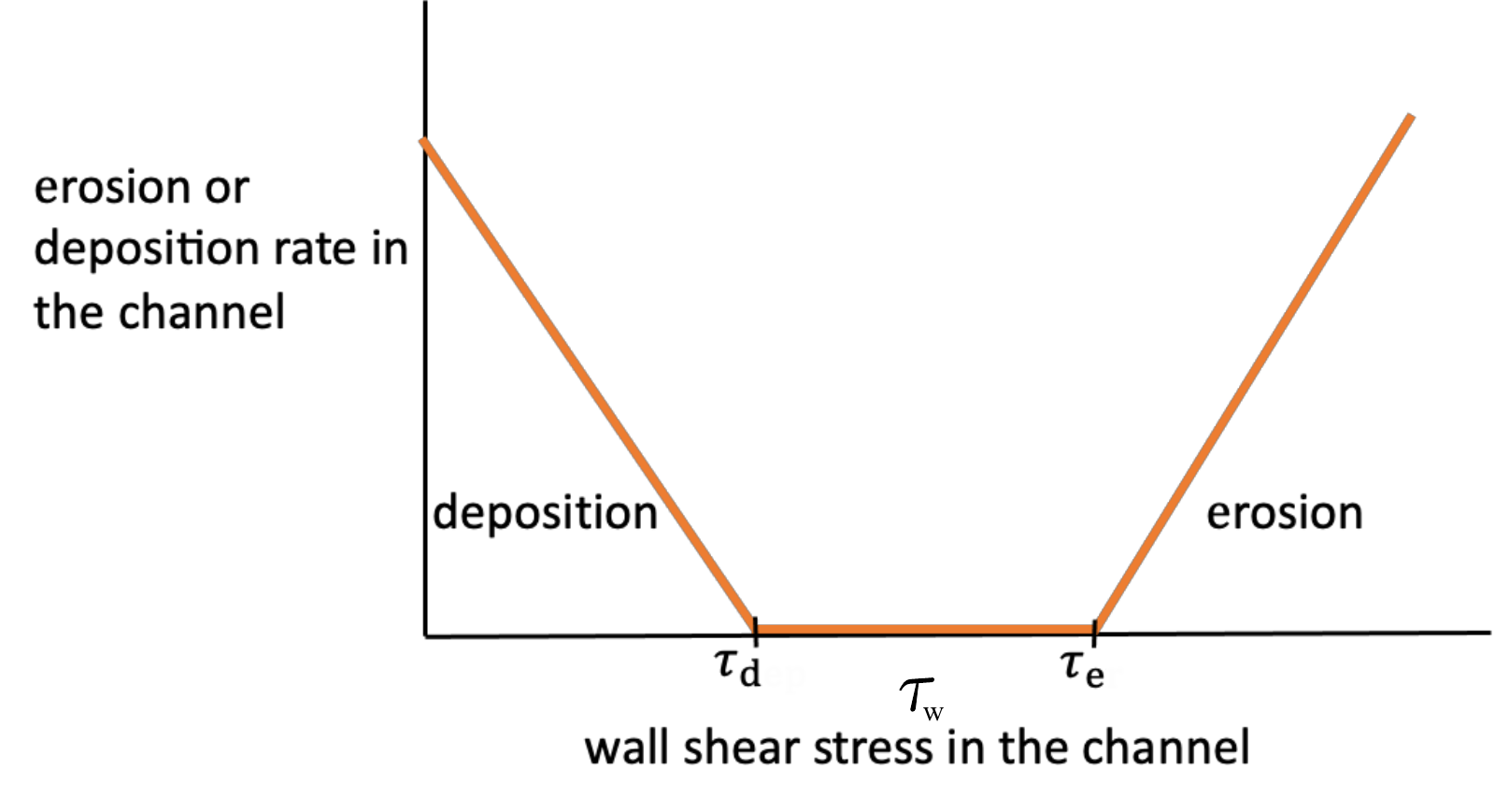}
    \caption{The deposition rate of the particles is highest at wall shear stress $\tau_w= 0$, and then monotonically decreases until $\tau_w = \tau_d$. The erosion rate of the deposited particles monotonically increases for $\tau_w >$ $\tau_e$. }
    \label{fig:tausketch}
\end{figure}

In summary, in the model, the deposition probability is highest at zero wall shear stress, and monotonically decreases until $\tau_w = \tau$ and then the erosion probability monotonically increases for $\tau_w > \tau $ as shown in the sketch FIG \ref{fig:tausketch}. After deposition, if the number of deposited particles in an edge grows larger than its capacity, the edge becomes blocked and temporarily removed from the network, and the pressure values and hydrodynamics are recalculated for the network. If the deposited particles are eroded in an edge such that it is no longer blocked, the network hydrodynamics are again calculated with the edge reinserted in the network. Throughout the simulation, we regularly apply the Kirchhoff's laws and solve for local pressure and flow rate values as the network changes.

\begin{center}
\begin{table*}
\caption{\label{tab:variables} The table contains a list of all the symbols used in this paper, what they denote, and their value or range of values from simulations in SI units. Similar values from past experiments are in \cite{BizmarkNavid2020Mdoc}.}
\begin{ruledtabular}
\begin{tabular}{ ccc  }
\hline
 \textbf{parameter} & \textbf{symbol} & \textbf{range}  \\ 
 \hline
 Pressure Difference Across the Device & $\Delta P$  & $6 \times 10^4 - 2.4 \times 10^5 Pa$ \\ 
 \hline
 Total Flow Rate & Q &   $10^{-12}-10^{-11} m^3/s$
  \\  
 \hline
 Local (Edge) Flow Rate & q & $10^{-14}-10^{-13} m^3/s$  \\  
 \hline
 Wall Shear Stress Threshold  & $\tau$ &   \\  
 \hline
 Wall Shear Stress & $\tau_w$ &   \\  
 \hline
 Hydraulic Resistance of the Porous Medium & $R$ &   \\  
 \hline
  Channel (Edge) Diameter & $d$ &   \\  
 \hline
Dynamic Viscosity & $\eta$ & $6 \times 10^{-2} Pa \cdot s $   \\  
 \hline
  Particle Diameter & $d_p$ & $1\mu m$  \\  
 \hline
   Erosion Coefficient \cite{JagerR2017Cipm} & $\kappa_e$ & $0.17 s/m$  \\  
 \hline
   Deposition Coefficient \cite{JagerR2017Cipm} & $\kappa_d$ & $1.7 s/m$  \\  
 \hline
\end{tabular}
\end{ruledtabular}
\end{table*}
\end{center}

\subsection*{S3. Categorizing the Localized Deposition Profiles} 
\label{s3:82}
\begin{figure}
    \centering
    \includegraphics[width =0.5 \textwidth]{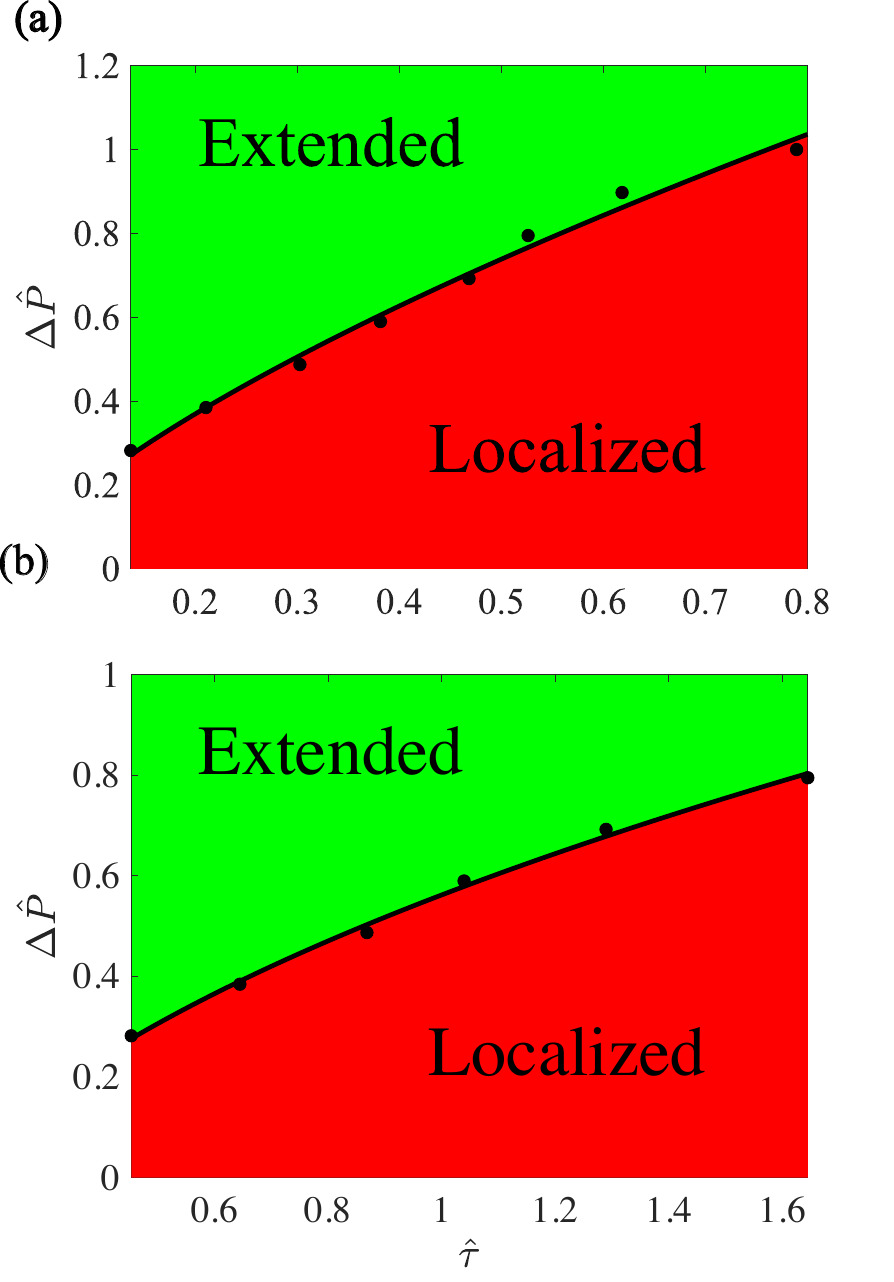}
    \caption{In the parameter space of normalized pressure $\Delta\hat{ P}$ and shear stress threshold $\hat{\tau}$, a boundary separates the two regimes of localized and extended deposition, reminiscent of the phases in \cite{MajumdarSatyaN1998Npti}. {In the Main Text we show a similar figure for an 82\% threshold for particles deposited in the first half of the medium.} Regardless of the specific categorization scheme, (a) choosing 75\% instead of 82\% for half-medium, or (b) tracking statistics for quarter-medium instead of half-medium and choosing 68\% as the transition point, we see the same essential results. The filled circles correspond to results gathered from 10 trials and the solid lines show the best fit curve of the form $ax^b+c$.}
    \label{fig:sm_phasediagram}
\end{figure}
We repeat the trials only varying the pressure difference and the threshold for erosion.  To quantitatively distinguish between localized and extended deposition, we consider the following definition for the former: by the end of the simulation, if the percentage of deposited particles in the first half of the medium is more than or equal to 82\%, we categorize the trial as localized. In other words, as we decrease the pressure difference while keeping other variables constant, we note the transition from extended to localized deposition as the first instance when 82\% of particles are deposited in the first half of the medium near the inlet. {To calculate these percentages, we use representative experimental data for each regime. From the images that show particle deposition across the medium, we calculate the percentage of pixels in the first half or first quarter of the medium, assuming that the number of deposited particles is proportional to the pixel count in images from the experiment.} We expect our essential results to be independent of this specific choice of the transition value.
\begin{center}
\begin{table}
\caption{\label{Tab:2} The table shows the normalized variables that appear in the Main Text and specific values used to normalize each variable. Multiplying each normalized variable by the value listed here would result in its physical value in SI units.}
\begin{ruledtabular}
\begin{tabular}{ cc  }
\hline
 \textbf{variable} & \textbf{normalization constant}  \\ 
 \hline
 ${\Delta \hat{P}}$  & $1.95 \times 10^5 Pa$ \\ 
 \hline
  $\hat{\tau}$ &  $3.8 Pa$ \\  
 \hline
 $\hat{L}$ & $6 \times 10^{-3} m$ \\
 \hline
\end{tabular}
\end{ruledtabular}
\end{table}
\end{center}
\begin{figure}
\begin{center}
   \includegraphics[width=0.5\textwidth]{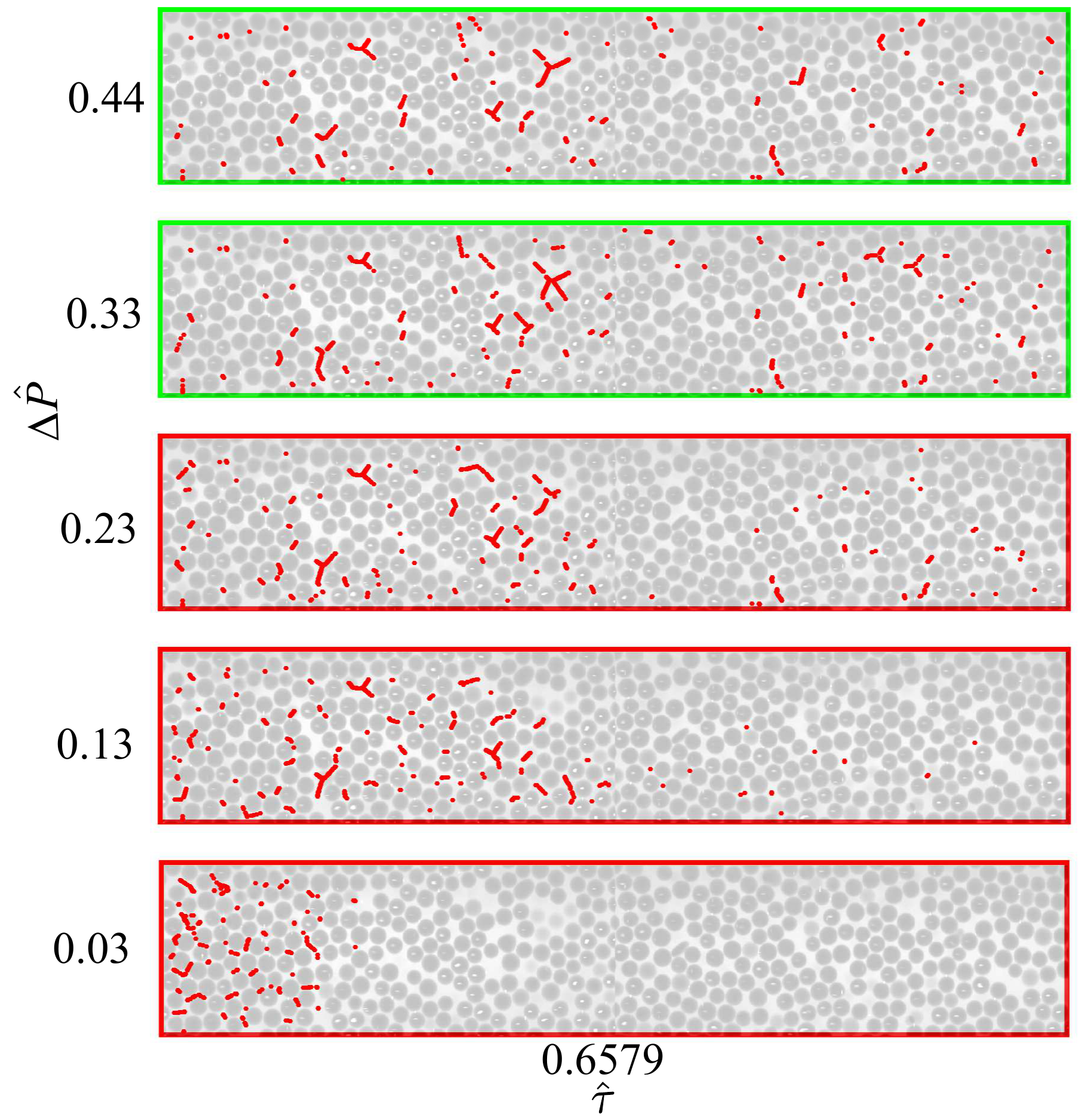}
\caption{Here, the size of the system is half the one used in the Main Text. The final frames of the simulation over a range of values of applied pressures $\Delta \hat{P}$ and at a representative shear threshold values $\hat{\tau}$ show a clear separation between the localized and extended deposition regimes similar to experimental observations and the results shown in the Main Text. The green or red border represents extended or localized, respectively.}
\label{fig:half}
\end{center}
\end{figure}
\begin{figure}
\begin{center}
   \includegraphics[width=0.5\textwidth]{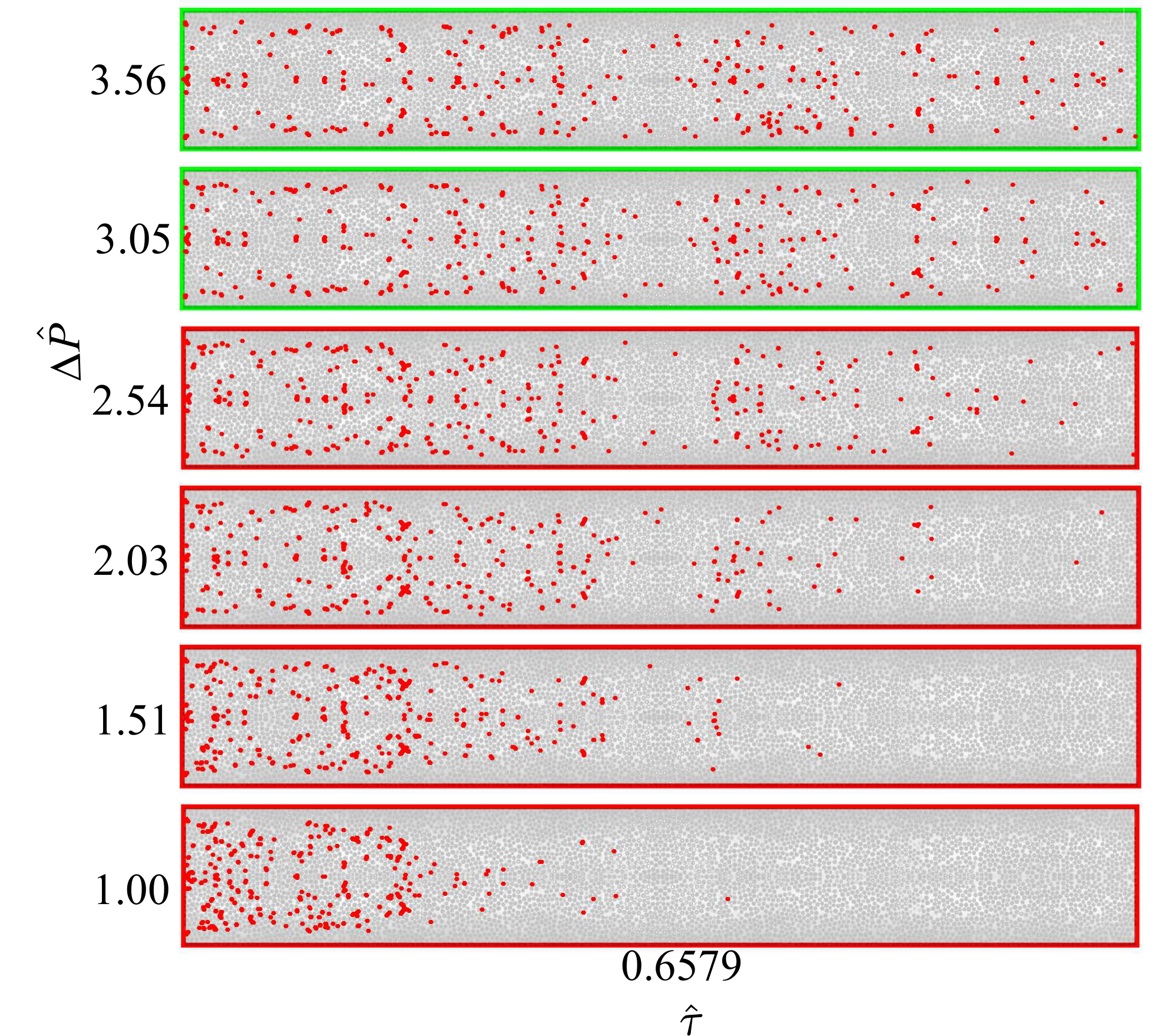}
\caption{Here, the size of the system is approximately double the one used in the Main Text. The final frames of the simulation over a range of values of applied pressures $\Delta \hat{P}$ and at a representative shear threshold values $\hat{\tau}$ show a clear separation between the localized and extended deposition regimes similar to experimental observations and the results shown in the Main Text. The green or red border represents extended or localized, respectively.}
\label{fig:dbl}
\end{center}
\end{figure}
To further test this assumption, we repeat the simulation trials for two different categorization schemes which are also supported by experimental data. In one case, corresponding to Fig. \ref{fig:sm_phasediagram} (a), if more than 75\% of the deposited particles were deposited in the first half of the medium, we categorize the trial as localized. In case two, corresponding to Fig. \ref{fig:sm_phasediagram} (b), if in one trial, more than 68\% of the deposited particles were deposited in the first quarter of the medium, we categorize that trial as localized. As Fig. \ref{fig:sm_phasediagram} demonstrates, regardless of the specific categorization, the essential results of the model remain the same as long as the choice of parameters are reasonable given the experimental observations. The critical transition values for pressure divide the phase space into two regions of localized and extended. For each case, we use the MATLAB curve fitting tool to find the best fit \cite{MATLAB}. {The fitted curves shown in Fig. \ref{fig:sm_phasediagram} are power laws of the form $ax^b+c$. These curves are shown as solid lines with $a = 1.28$, $b = 0.67$, and $c = -0.067$ in Fig. \ref{fig:sm_phasediagram} (a) and $a = 0.93, b = 0.46$, and $ c = -0.37$ in Fig. \ref{fig:sm_phasediagram} (b) respectively.}  The search for the transition point involves repeating the trials at least 30 times for each set of $\hat{\tau}$ and $\Delta \hat{P}$, 10 for the transition point with an uncertainty of $0.5\%$, 10 for a lower bound, 10 for an upper bound.

{Past studies on colloidal particle interactions similarly show that varying the global hydrodynamics such as applied pressure or flow rate lead to different deposition regimes consistent with our observations shown in FIG. \ref{fig:phasecurve} and FIG. \ref{fig:sm_phasediagram}. In particular, in \cite{BoekMeso2008} Boek et al.~simulate colloidal flow of asphaltene particles and find that increasing the interaction strength leads to different deposition regimes where permanent clogging occurs for stronger interactions. Note that, however, this is not directly comparable to our work since the channel considered in these models corresponds to one of the many channels in our simulations. Their work considers a more detailed simulation of deposition dynamics in a single channel while our work considers a network of such channels with phenomenological laws for the erosion and deposition.}

One difference between the experiments and simulations is in the chronological progression of the deposition profile. In the experiments, the deposition profile tends to become more localized over time. In contrast, as shown in FIG. \ref{fig:Fx}, in simulations, the deposition profile becomes more uniform at later times. One possible explanation lies in the details of clogging mechanism. In experiments, as more particles aggregate at the inlet, the mechanism behind clogging of particles evolves while on the contrary, the simulations assume a simple clogging mechanism and do not capture the details of clogging such as cake formation \cite{cakeTienChi2014Abmo, cakeHwangKuo-Jen2002Cmoa} due to coarse-graining and 2D consideration of this 3D problem.

All results presented in the Main Text correspond to one bead packing. Our algorithm, however, works for other bead packing networks, as well. Here, we demonstrate that the transition occurs in systems of different sizes (taking the size to be half and double). As FIG. \ref{fig:half} and FIG. \ref{fig:dbl} show, starting from the localized regime and at a set shear stress threshold, increasing the pressure leads to a more uniform deposition and eventually, to the extended regime. Here, we have used the 82\% threshold to categorize extended and localized cases similar to the Main Text. Notably, as FIG. \ref{fig:dbl} shows the algorithm works well for larger networks in which the number of edges is several times greater than in the original network.






\end{document}